\documentclass[12pt]{article}
\usepackage{amsfonts,amsmath,epsfig,graphicx,amssymb}
\usepackage{amsmath,epsfig,graphicx,amssymb}
\usepackage[margin=1in]{geometry}
\usepackage{amsmath}
\usepackage{amsfonts}
\usepackage{amssymb}
\usepackage{makeidx}
\usepackage{graphicx}
\usepackage{float}
\newcommand{\beq}{\begin{equation}}
\newcommand{\eeq}{\end{equation}}
\newcommand{\ben}{\begin{eqnarray}}
\newcommand{\een}{\end{eqnarray}}
\date{}
\begin{document}
\title{Intersystem Non-separability and CHSH-Bell Violations in Classical Optics}
\author{Partha Ghose\footnote{partha.ghose@gmail.com} \\
Tagore Centre for Natural Sciences and Philosophy, Kolkata, India}
\maketitle
\begin{abstract}
A method is proposed to produce a classical optical state that is `intersystem nonseparable' and a close analog of the $\phi^+$ Bell state. A derivation of the CHSH-Bell inequality is sketched within the framework of classical polarization optics, using non-contextuality for factorizable states as an axiom rather than any hidden variable theory. It is shown that the classical state violates this inequality. 
\end{abstract}
\section{Introduction}
There is some controversy in the literature as to whether the nonseparability of different degrees of freedom that has been observed in a classical light beam should be called `entanglement' because `entanglement' is a term coined by Schr\"{o}dinger specifically to express a purely quantum nonseparability that persists in spite of arbitrary spatial separation of the sub-systems \cite{boyd} (2019). It is true that most instances of nonseparability observed with classical light so far are of a local nature, involving different modes (DoFs) of the same optical beam \cite{qian1, qian2}. They have been termed `intra-system nonseparability' or 'classical entanglement' as opposed to 'quantum entanglement' which is believed (by the majority of physicists) to imply nonlocality \cite{ai}. It is also generally believed that only intrasystem nonseparability has a classical analogy, though there is no proof. 

It is therefore important to investigate whether `intersystem nonseparability' also has any classical analog, and if so, what it implies. In this paper I will first propose a method of producing an intersystem nonseparable state in classical polarization optics, and then explore to what extent it is quantum-like.

To describe polarized light it is most convenient to use normalized Jones vectors
\beq
 \vert J\rangle = \frac{1}{\sqrt{\langle J\vert J\rangle}}\left(\begin{array}{c}
 E_x\\ E_y
\end{array} \right)
\eeq
with $E_x = A_0 \hat{e}_x {\rm exp (i\phi_x)}$ and $E_y = A_0 \hat{e}_y {\rm exp (i\phi_y)}$ the complex transverse electric fields, $\hat{e}_x$ and $\hat{e}_y$ the unit polarization vectors, and $\langle J\vert J\rangle = \vert E_x\vert^2 + \vert E_y\vert^2 =  A_0^2$, the intensity $I_0$.
This is a column vector in a 2-dimensional Hilbert space of polarization states. The Jones vector for horizontally ($x$ axis) polarized light is
\beq
\vert h\rangle \equiv \vert 1\rangle = \left(\begin{array}{c}
 1\\ 0
\end{array}\right),
\eeq
and that of vertically polarized ($y$ axis) light is
\beq
\vert v\rangle\equiv \vert 0\rangle = \left(\begin{array}{c}
 0\\ 1
\end{array}\right).
\eeq
Jones vectors for other linear polarization states can be written as linear combinations of these basis vectors. 
\section{How to Produce a Classical Bell-like State}
Let us now briefly recollect the standard procedure for producing the first Bell state in quantum mechanics. One uses a two-qubit circuit consisting of a Hadamard gate H and a CNOT gate. The Hadamard gate applied to a general qubit gives 
$H|q\rangle = \frac{1}{\sqrt{2}}[|0\rangle + |1\rangle]$. This state combined with a $|0\rangle$ state produces the state $\frac{1}{\sqrt{2}}[|0\rangle \otimes |0\rangle + |1\rangle \otimes |0\rangle]$, and the CNOT gate produces the Bell state $|\phi^+\rangle =\frac{1}{\sqrt{2}}[|0\rangle \otimes |0\rangle + |1\rangle \otimes |1\rangle]$.
\begin{figure}[H]
\centering
{\includegraphics[scale=0.5]{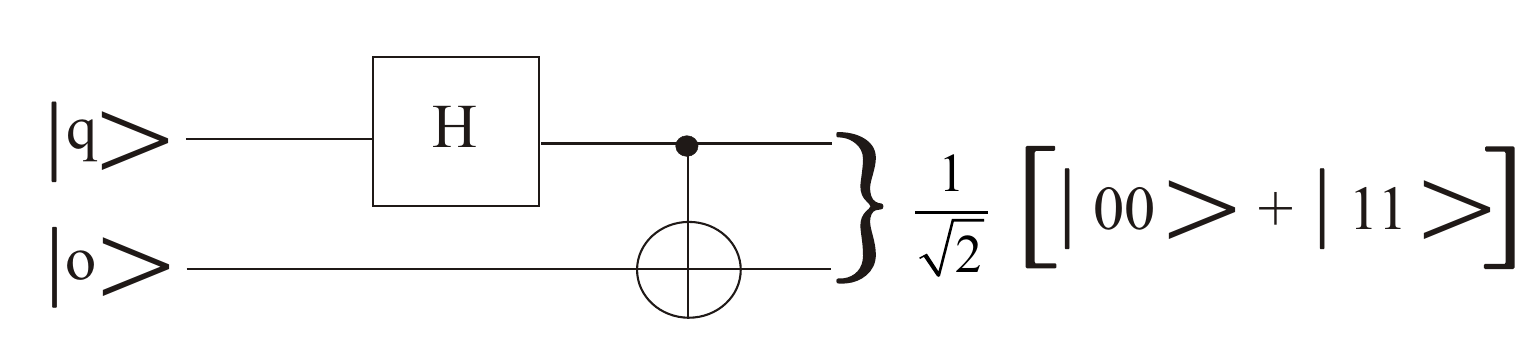}}
\caption{\label{Figure 1.1}{\footnotesize Circuit diagram for production of the Bell state $|\phi^+\rangle$. $H$ is a Hadamard gate and the circle in the second line a CNOT gate.}}
\end{figure}

Let us now see how this circuit can be realized in classical optics. 

\begin{figure}[H]
\centering
{\includegraphics[scale=0.5]{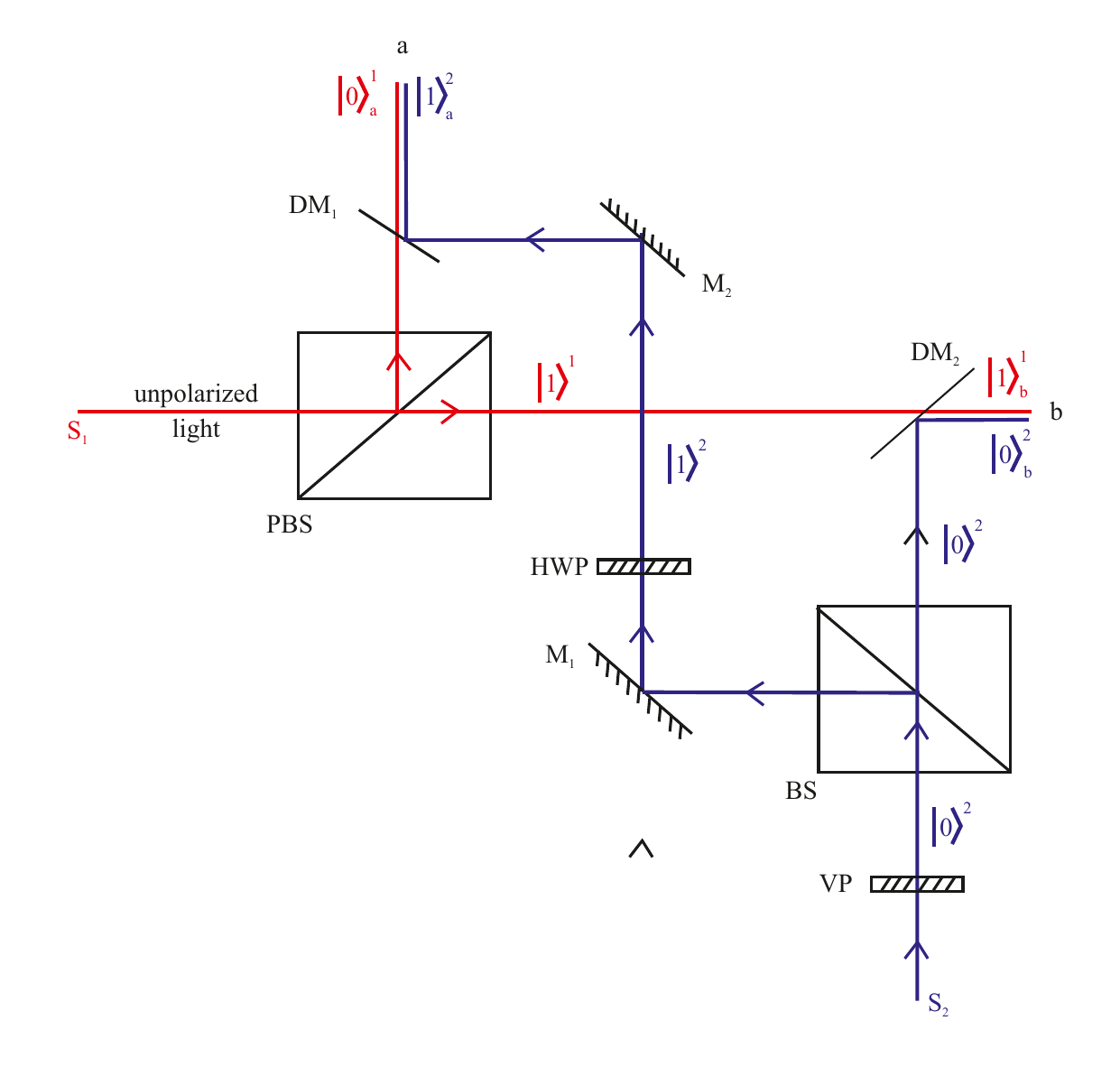}}
\caption{\label{Figure 1.2}{\footnotesize Schematic diagram to produce a classical analog of the first Bell state.}}
\end{figure}
 
Consider two quasi-monochromatic sources $S_1$ and $S_2$ of classical light of equal intensities but different mean angular frequencies $\omega_1$ (denoted by red in the figure) and $\omega_2$ (denoted by blue in the figure). Let the beam emitted by $S_1$ be unpolarized and that emitted by $S_2$, also unpolarized, be converted to a vertically polarized beam $|0\rangle_2$ by passing through a vertical polarizer VP. These correspond to the qubit $|q\rangle$ and $|0\rangle$. Let the unpolarized beam (red) be incident on a polarizing beam splitter PBS as shown in the figure. It acts like a Hadamard gate, producing the polarization state $\frac{1}{\sqrt{2}}[i|0\rangle_1^a + |1\rangle_1^b]$, the reflected light being vertically polarized ($|0\rangle$) and the transmitted light horizontally polarized ($|1\rangle$). Let the second beam (vertically polarized $|0\rangle_2$ and blue) be incident on an ordinary beam splitter BS. The part of the beam transmitted by BS is combined with the horizontally polarized beam transmitted by PBS by means of a dichroic mirror $DM_2$. The part of the beam reflected by BS is first reflected by a lossless mirror $M_1$, passes throgh a half-wave plate HWP, gets converted into a horizontally polarized beam $|1\rangle_2$ which is then reflected by a lossless mirror $M_2$ and combined with the vertically polarized red beam reflected by PBS by means of a dichroic mirror $DM_1$.The beam splitter BS and the half-wave plate HWP thus act like a CNOT gate, and the final polarization state of the output beams $a$ and $b$ is
\beq
|\phi^+\rangle = \frac{i}{\sqrt{2}}\left[|0\rangle_1^a\otimes|0\rangle_2^b + |1\rangle_1^b\otimes|1\rangle_2^a\right]. \label{nonsep}
\eeq
This is a nonseparable/nonfactorizable state. It is precisely what we were looking for, namely `intersystem nonseparability' of classical light, a nonseparability that persists in spite of arbitrary spatial separation of the sub-systems. The three other Bell-like nonseparable states can be produced by using variations of this method.

Let us see to what extent such states are quantum-like. Shimony \cite{shim} has listed the following three criteria that any system S 
produced for testing its quantum nature must satisfy:

(I) In any state of a physical system S there are some eventualities which have indefinite
truth values.

(II) If an operation is performed which forces an eventuality with indefinite truth value to
achieve definiteness ... the outcome is a matter of chance.

(III) There are `entangled systems' (in Schr\"{o}dinger's phrase) which have the property that
they constitute a composite system in a pure state, while neither of them separately is in a
pure state. 

The state (\ref{nonsep}) clearly satifies criterion (III). As regards the other two criteria, it would be useful to consider thermal light. In particular, let $S_1$ and $S_2$ be two thermal sources which are wide-sense stationary. In a beam of thermal light the value of the `optical field' is unpredictable. The observable optical field of such a light beam is the space-time dependent electric field vector
\beq
\vec{E}(r, t) = \hat{h} E_h(r, t) + \hat{v} E_v(r, t) 
\eeq
where $\hat{h}$ and $\hat{v}$ are orthogonal and arbitrarily oriented `horizontal' and `vertical' polarization directions, and the amplitudes $E_h(r, t) = \sum_n h_n(t)\phi_n(r)$ and $E_v(r, t) = \sum_l v_l(t)\phi_l(r)$ are statistically completely uncorrelated, being elements of a stochastic function space. Scalar products of the vectors in this space correspond physically to observable correlation functions such as $\langle E_v E_h\rangle$, the angular bracket denoting expectation value. For unpolarized light $\langle E_v E_v\rangle = \langle E_h E_h\rangle$ and $\langle E_v E_h\rangle = 0$. The $\phi_n(r)$ are orthonormal spatial mode functions with $\int dr \phi^*_ n(r)\phi_l(r) = \delta_{n,l}$, and $h_n(t)$ and $v_l(t)$ represent stochastic random coefficients whose origin one can assign to distant and unknowable dipole radiators. 

Since the value of $\vec{E}$ in thermal light is unpredictable at any time, and a definite truth value is obtained only when the field is appropriately detected, Shimony's criteria (I) and (II) are also satisfied. Nevertheless, this picture of the optical field does not imply that it is quantized, only a non-deterministic viewpoint is employed to extract predictions from the randomly unknown ensemble of field potentialities. 

Since the red beam with frequency $\omega_1$ is unpolarized, 
\ben
\langle E_{1v}(r,t), E_{1h}(r,t)\rangle &=& 0,\label{a}\\
\langle E^*_{1v}(r,t), E_{1v}(r,t)\rangle &=& \langle E^*_{1h}(r,t), E_{1h}(r,t)\rangle \label{b}
\een
for all $(r,t)$ \cite{qian3}. Since this unpolarized red beam splits into vertical and horizontal components at some point within the beam splitter PBS (as shown in Fig. 1), the first equation ($\ref{a}$) implies that the red beams in paths $a$ and $b$ are completely uncorrelated. In other words, {\em the red beam will exhibit anticoincidence on the beam splitter, a quantum-like feature}. The second result ($\ref{b}$) implies that the measured intensities of the vertical and horizontal components are always equal everywhere in the beams. 

Now, the classical electrical fields in the two paths $a$ and $b$ (Fig. 1) are (see Appendix)
\ben
\vec{E}^a(r_a,t) &=& \hat{v} E^a_{1v}(r_a, t) + \hat{h} E^a_{2h}(r_a, t),\label{Ea}\\
\vec{E}^b(r_b,t) &=& \hat{h} E^b_{1h}(r_b, t) + \hat{v} E^b_{2v}(r_b, t).\label{Eb}
\een
They fluctuate stationarily in the wide-sense. Since the sources $S_1$ and $S_2$ are independent, the following first-order cross correlations vanish, namely
\ben
\langle E^{a*}_{1v}(r_a,t), E^a_{2h}(r_a, t + \tau)\rangle &=& \langle E^{b*}_{2v}(r_b,t), E^b_{1h}(r_b,t + \tau)\rangle = 0,\label{cc1}\\
\langle E^{a*}_{1v}(r_a,t), E^b_{2v}(r_b, t)\rangle &=& \langle E^{a*}_{2h}(r_a,t), E^b_{1h}(r_b,t)\rangle = 0 \label{cc2}.
\een
Eqn (\ref{cc1}) implies that red and blue lights (which, by experimental design, always have orthogonal polarizations in each path) are completely uncorrelated in each path. {\em These results guarantee randomness of outcomes (red/blue) in each path/end, another quantum-like feature}. 
Eqn (\ref{cc2}) shows that the vertically polarized red beam in path $a$ and the vertically polarized blue beam in path $b$ are uncorrelated, and so also are the horizontally polarized red beam in path $b$ and the horizontally polarized blue beam in path $a$, in spite of the light being in the nonseparable polarization state (\ref{nonsep}). {\em This guarantees that `superluminal signalling' does not occur at the statistical level, which is also a quantum-like feature}.

These results demonstrate to what extent intersystem nonseparability in classical optics is similar to that in quantum optics. 

\section{CHSH-Bell Inequality}
The question that arises therefore is: does intersystem nonseparability of classical light imply any Bell violation which is usually taken as a measure of nonlocality? To see that, consider a classical separable (i.e. factorizable) state like
\beq
\frac{1}{\sqrt{2}} \left[|1\rangle_1^a\otimes|0\rangle_2^b + |0\rangle_1^a\otimes|0\rangle_2^b\right] = \left[|1\rangle_1^a + |0\rangle_1^a\right]\otimes |0\rangle_2^b.  \label{sep}
\eeq
If one treats the normalized classical Jones vectors mathematically as qubits, the state (\ref{sep}) is mathematically isomorphic to a separable two-particle state. Now, consider (a) joint polarization measurements on the two sides $a$ and $b$ at angles $(\theta_a, \theta_{a^\prime})$ and $(\theta_b, \theta_{b^\prime})$, and (b) instead of local realism (i.e. the requirement that measurements on one side cannot instantaneously affect the state on the other side), assume {\em noncontextuality} (which is the requirement that the result of a measurement is predetermined and not affected by how the value is measured, i.e. not affected by previous or simultaneous measurement of any other compatible or co-measureable observable). Mathematically the two assumptions are equivalent, only the interpretations are different. Thus it follows immediately from well established results \cite{chsh, bell} that the inequality  
\begin{equation}
-2 \leq E(\theta_a,\theta_b) - E(\theta_a,\theta_{b^\prime}) + E(\theta_{a^\prime}, \theta_b) + E(\theta_{a^\prime}, \theta_{b^\prime}) \leq +2
\end{equation}
must hold, where $E(\theta_a, \theta_b)$ are expectation values. [The outcomes need not be $\pm 1$ as is usually assumed. It is sufficient that they lie between $\pm 1$ which is guaranteed by Malus' Law in classical polarization optics with normalized Jones vectors \cite{ghose}.] If the nonseparable state (\ref{nonsep}) is used, one gets
\begin{equation}
E(\theta_a, \theta_b) = \cos (\theta_a - \theta_b),
\end{equation}
and for a specific set of values $\theta_a = \pi/2, \theta_{a^\prime} = 0, \theta_b = \pi/4, \theta_{b^\prime} = - \pi/4$ one gets $S = 2\sqrt{2}$ which violates the CHSH inequality. For a more detailed account of CHSH violations with noncontextuality, see the Appendix II.

The original derivations of the Bell and CHSH inequalities involved the use of local hidden variables (local realism). Hence, in such a framework a CHSH violation would rule out local hidden variable theories. If the hidden variables are noncontextual (Einstein realism), nonlocality would follow. To avoid nonlocality, one must assume a reality that is contextual (as Bohr did in terms of his complementarity idea \cite{bohr}).

As we have just seen, Bell-CHSH inequalities can also be derived within the framework of classical polarization optics using {\em noncontextuality}. Hence, a CHSH violation with spatially separated subsystems in classical polarization optics would imply a violation of noncontextuality rather than of locality \cite{khren1, khren2}, signalling a radical change in the very notion of classical realism. The entire edifice of classical physics (classical mechanics and classical field theory) has so far stood on the concept of a reality that is noncontextual, i.e. independent of observations, observations only revealing pre-existing values. The discovery of nonseparability in classical optics has changed this, revealing that even in the classical domain observables cannot be thought of simply as revealing pre-existing values.

\section{Concluding Remarks}

I have proposed a method to produce a classical optical state that is (a) `intersystem nonseparable', (b) statistically close to a Bell state and (c) violates the Bell-CHSH inequality without violating locality. Hence, intersystem nonseparability {\em per se} does not imply nonlocality.

Having said that, it is important to point out a crucial difference between classical and quantum nonseparable states, namely that the statistics can never be sub-Poissonian with classical light. 

\section{Appendix I}
To see that the electric fields (\ref{Ea}) and (\ref{Eb}) indeed represent the polarization state (\ref{nonsep}), consider the corresponding states
\ben
|{\cal{E}}^a\rangle &=& e^a_{1v}|0\rangle^a_1 + e^a_{2h}|1\rangle^a_2 ,\\
|{\cal{E}}^b\rangle &=&  e^b_{1h}|1\rangle^b_1 + e^b_{2v}|0\rangle^b_2,
\een
where $e^a_{1v}$ etc are the normalized electric field components.
Then, the normalized tensor product of the two independent light beams is 
\beq
\frac{1}{\sqrt{2}}\left[ e^a_{1v}e^b_{2v}|0\rangle^a_1\otimes |0\rangle^b_2 + e^b_{1h}e^a_{2h}|1\rangle^b_1\otimes |1\rangle^a_2\right].  
\eeq
The normalized electric field coefficients are usually omitted in writing Bell-like states.
 
To produce the nonseparable polarization state
\beq
|\psi^+\rangle = \frac{i}{\sqrt{2}}\left[|0\rangle_1^a\otimes|1\rangle_2^b + |1\rangle_1^b\otimes|0\rangle_2^a\right], \label{nonsep2}
\eeq
the experimental set up needs to be a variant of the one proposed here, namely one using the scheme $H|q\rangle\otimes |1\rangle  = \frac{1}{\sqrt{2}}[|0\rangle + |1\rangle]\otimes |1\rangle = \frac{1}{\sqrt{2}}[|0\rangle\otimes |1\rangle + |1\rangle \otimes |1\rangle]$ followed by a CNOT gate. This means replacing the vertical polarizer VP in the path of the $S_2$ beam (Fig. 1) by a horizontal polarizer VH. Then the electrical fields in the two paths are
\ben
\vec{E}^a(r_a,t) &=& \hat{v} E^a_{1v}(r_a, t) + \hat{v} E^a_{2v}(r_a, t),\\
\vec{E}^b(r_b,t) &=& \hat{h} E^b_{1h}(r_b, t) + \hat{h} E^b_{2h}(r_b, t),
\een
corresponding to the polarization state (\ref{nonsep2}).

\section{Appendix II}

To derive a Bell-CHSH inequality based on noncontextuality, let us first consider an arbitrary general state
\begin{eqnarray}
|\Phi\rangle &=& (\cos\alpha|0\rangle_1^a + e^{i\beta}\sin\alpha|1\rangle_2^a)(\cos\gamma|0\rangle_2^b + e^{i\delta}\sin\gamma|1\rangle_2^b)\\
&=& |\psi\rangle_a|\psi\rangle_b\label{gen}
\end{eqnarray}
where $\alpha, \beta, \gamma, \delta$ are arbitrary parameters. Next, let us define the
correlation
\begin{eqnarray}
E(\theta,\phi)=\langle \Phi\vert \sigma_{\theta}.\sigma_{\phi}\vert \Phi\rangle
\end{eqnarray}
where
\begin{eqnarray}
\sigma_{\theta}=\sigma_{\theta,0}-\sigma_{\theta,\pi},\\
\sigma_{\phi}=\sigma_{\phi,0}-\sigma_{\phi,\pi},
\end{eqnarray}
with
\begin{eqnarray}
\sigma_{\theta,0}=\frac{1}{2}(|0\rangle_1^a + e^{i\theta}|1\rangle_2^a)(\langle 0|_1^a + e^{-i\theta}\langle 1|_2^a) \otimes \mathbb{I}_b,\nonumber \\
\sigma_{\theta,\pi}=\frac{1}{2}(|0\rangle_1^a - e^{i\theta}|1\rangle_2^a)(\langle 0|_1^a - e^{-i\theta} \langle 1|_2^a)\otimes \mathbb{I}_b,\nonumber \\
\sigma_{\phi,0}= \mathbb{I}_a\otimes\frac{1}{2}(|0\rangle_2^b + e^{i\phi}|1\rangle_2^b)(\langle 0|_2^b + e^{-i\phi}\langle 1|_2^b),\nonumber \\
\sigma_{\phi,\pi}= \mathbb{I}_a\otimes\frac{1}{2}(|0\rangle_2^b - e^{i\phi}|(1\rangle_2^b)(\langle 0|_2^b - e^{-i\phi}\langle 1|_2^b),\label{sigma}
\end{eqnarray}
where $\theta$ and $\phi$ are phase shifts between the two polarization modes in the $a$ and $b$ beams respectively, and $\mathbb{I}_b$ and $\mathbb{I}_a$ are the identity operators in the Hilbert space $\mathcal{H}_b$ spanned by the $b$ beam and the Hilbert space $\mathcal{H}_a$ spanned by the $a$ beam respectively.
Hence,
\begin{eqnarray}
\sigma_{\theta}=\left(e^{-i\theta}|0\rangle_1^a\langle 1|_2^a + e^{i\theta}|1\rangle_2^a \langle 0|_1^a \right)\otimes \mathbb{I}_b,\\
\sigma_{\phi}= \mathbb{I}_a\otimes \left(e^{-i\phi}|0\rangle_2^b\langle 1|_2^b + e^{i\phi}|1\rangle_2^b\langle 0|_2^b \right).
\end{eqnarray} 
It should be noted that $\sigma_{\theta}$ and $\sigma_{\phi}$ act upon different Hilbert spaces altogether, and hence they commute with each other. This property is necessary for noncontextuality of the total system. It has always been a tenet of classical physics that whatever exists in the physical world is independent of observations which only serve to reveal them. Put more technically, this means that the result of a measurement is predetermined and is not affected by how the value is measured, i.e. not affected by previous or simultaneous measurement of any other {\em compatible} or co-measureable observable. Hence the need for commuting observables to test noncontextuality.

For the general product state (\ref{gen}),
\begin{eqnarray}
E(\theta,\phi)&=&\langle\psi_a|\sigma_{\theta}|\psi_a\rangle\langle\psi_b|\sigma_{\phi}|\psi_b\rangle\nonumber \\
&=& E_a(\theta)E_b(\phi),\label{exp}
\end{eqnarray}
with
\begin{eqnarray}
E_a(\theta)=\sin 2\alpha\cos(\beta-\theta),\label{epol}\\
E_b(\phi)=\sin 2\gamma\cos(\delta-\phi).\label{epath}
\end{eqnarray}
Thus, the expectation value $E(\theta,\phi)$ is the product of the expectation values $E_a(\theta)$ and $E_b(\phi)$. Hence, for product states in classical optics, the polarization measurements on the two beams are independent of one another in all contexts (i.e. for all parameter settings). This is the content of {\em noncontextuality}. This may, at first sight, look obvious and trivial, but on closer inspection, one finds that it implies the inequality
\beq
-1\leqslant E(\theta,\phi)\leqslant 1 \label{cor}
\eeq
for the correlation.

Now, define a quantity $S$ as
\ben
S(\theta_{1},\phi_{1};\theta_{2},\phi_{2}) &=& E(\theta_{1},\phi_{1})+E(\theta_{1},\phi_{2})-E(\theta_{2},\phi_{1})+E(\theta_{2},\phi_{2}).
\een
It follows from (\ref{cor}) that
\beq
|S|\leq 2.\label{bound}
\eeq
This is the CHSH inequality. All that is required to derive this bound for product states is that {\em the correlations lie between $-1$ and $+1$}, which is guaranteed by the results (\ref{epol}) and (\ref{epath}). 
 
Consider now the normalized Bell state 
\beq
|\phi^+\rangle = \frac{i}{\sqrt{2}}\left[|0\rangle_1^a\otimes|0\rangle_2^b + |1\rangle_1^b\otimes|1\rangle_2^a\right]. \label{ent}
\eeq
The correlation calculated for such a state is
\ben
E(\theta,\phi)&=&\langle \phi^+|\sigma_{\theta}\cdot\sigma_{\phi}|\phi^+\rangle\nonumber\\
&&\:= \langle \phi^+\vert \,[(+)\sigma_{\theta,0} + (-)\sigma_{\theta,\pi} ]. [(+)\sigma_{\phi,0} + (-)\sigma_{\phi,\pi} ]\vert \phi^+\rangle\\
&=&\langle \phi^+[\sigma_{\theta,0}\cdot\sigma_{\phi,0}+\sigma_{\theta,\pi}\cdot\sigma_{\phi,\pi}\nonumber\\ &&\: -\sigma_{\theta,0}\cdot\sigma_{\phi,\pi}-\sigma_{\theta,\pi}\cdot\sigma_{\phi,0}]\vert \phi^+\rangle
\een
The intensities corresponding to the four possible orientations are given by
\ben
I(\theta,\phi)&=& \langle \phi^+|\sigma_{\theta,0}\cdot\sigma_{\phi,0}\vert \phi^+\rangle,\nonumber\\
I(\theta+\pi,\phi+\pi)&=& \langle \phi^+|\sigma_{\theta,\pi}\cdot\sigma_{\phi,\pi}\vert \phi^+\rangle,\nonumber\\
I(\theta+\pi,\phi)&=& \langle \phi^+|\sigma_{\theta,\pi}\cdot\sigma_{\phi,0}\vert \phi^+\rangle,\nonumber\\
I(\theta,\phi+\pi)&=& \langle \phi^+|\sigma_{\theta,0}\cdot\sigma_{\phi,\pi}\vert \phi^+\rangle,
\een
where clearly $I(\theta,\phi)=\frac{1}{2}[1+\cos(\theta+\phi)]$ from (\ref{ent}) and the definitions (\ref{sigma}). 

We can write $E(\theta, \phi)$ in terms of the normalized intensities as
\ben
E(\theta,\phi)&=&\dfrac{I(\theta,\phi)+I(\theta+\pi,\phi+\pi)-I(\theta+\pi,\phi)-I(\theta,\phi+\pi)}{I(\theta,\phi)+I(\theta+\pi,\phi+\pi)+I(\theta+\pi,\phi)+I(\theta,\phi+\pi)}\nonumber\\
&=& {\rm cos}(\theta + \phi).
\een
It is clear from this that the noncontextuality bound (\ref{bound}) is violated by the state (\ref{ent}) for the set of parameters $\theta_1 = 0, \theta_2 = \pi/2, \phi_1 = \pi/4, \phi_2 = - \pi/4$ for which $\vert S\vert = 2\sqrt{2}$. 

\section{Acknowledgement}
The author is grateful to C. S. Unnikrishnan, G. Raghavan and members of the School of Quantum Technologies, DIAT, Pune for helpful discussions on the difference between quantum and classical entanglement. I am also grateful to the helpful comments received from the Research Feedback team to improve the presentation.


\begin{thebibliography}{0}
\bibitem{boyd}
D. Paneru, E. Cohen, R. Fickler, R. W. Boyd and E. Karimi1, arXiv:1911.02201 [quant-ph] (2019) and references therein
\bibitem{qian1}
X-F Qian, B. Little, J. C. Howell and J. H. Eberly, `Violation of Bell's Inequalities with Classical Shimony-Wolf States: Theory and Experiment', 	arXiv:1406.3338 [quant-ph] (2014).
\bibitem{qian2}
X-F Qian, B. Little, J. C. Howell, AND J. H. Eberly, {\\bf 2} (7), 611-615 (2015).
\bibitem{ai}
A. Aiello, F. T\"{o}ppel1, C. Marquardt, E. Giacobino and G. Leuchs, {\em New J. of Phys.} {\bf 17}, 043024 (2015).
\bibitem{shim}
A. Shimony, {\em Brit. J. Phil. Sci.} {\bf 35}, 25-45 (1984).
\bibitem{qian3}
X-F Qian, B. Little, J. C. Howell, and J. H. Eberly, {\em Optica} {\bf 2} (7), 611-615 (2015).
\bibitem{bross}
C. Brosseau, {\em Fundamentals of Polarized Light: A Statistical Optics Approach}, Wiley, New York (1998).
\bibitem{wolf}
E. Wolf, {\em Introduction to the Theory of Coherence and Polarization of Light}, Cambridge Univ. Press (2007).
\bibitem{chsh}
J. F. Clauser, M. A. Horne, A. Shimony and R. A. Holt, {\em Phys. Rev. Lett.} {\bf 23}, 880-884, (1969).
\bibitem{bell}
J.S. Bell, {\em Physics} {\bf 1} (3), 195-200 (1964), reproduced as Ch. 2 of J. S. Bell, {\em Speakable and Unspeakable in Quantum Mechanics}, Cambridge University Press (2004).
\bibitem{ghose}
P. Ghose and A. Mukherjee, {\em Adv. Sc., Eng. and Med.} {\bf 6}, 246-251 (2014).
\bibitem{bohr}
N. Bohr, {\em Phys. Rev.} {\bf 48}, 696-702 (1935).
\bibitem{khren1}
A. Khrennikov, Quantum versus classical entanglement: eliminating
the issue of quantum nonlocality. arXiv:1909.00267v1 [quant-ph].
\bibitem{khren2}
A. Khrennikov, Two faced Janus of quantum nonlocality. arXiv: 2001.02977v1 [quant-ph].
\end{thebibliography}
\end{document}